\documentclass[reprint,amsmath,amssymb,aps]{revtex4-1}
\usepackage{graphicx}

\begin{document}

\title{Computation of products of phase space factors and nuclear matrix elements for the Double Beta Decay\thanks{Supported by Ministry of Research and Innovation through UEFISCDI, project PCE-2016-0078, contract 198/2017 }}

\author{%
	S. Stoica$^{(1,2)}$ 
}

\affiliation{%
$^1$ International Centre for Advanced Training and Research in Physics, Magurele 077125, Romania\\
$^2$ Horia Hulubei National Institute of Physics and Nuclear Engineering, Magurele 077125, Romania\\
}
\email{sabin.stoica@cifra.infim.ro}

\begin{abstract}

The  nuclear matrix elements (NME) and phase space factors (PSF) entering the half-life formulas of the double-beta decay (DBD) process are two key quantities whose accurate computation still  represents a challenge. In this paper we propose a new approach of calculating them, namely to compute directly their product as an unique formula. This procedure allows a more coherent treatment of the nuclear approximations and input parameters appearing in both quantities and avoids possible confusion in interpreting the DBD data due to different individual expressions adopted for PSF and NME (and consequently their reporting in different units) by different authors. Our calculations are performed for both two neutrino ($2\nu\beta\beta$) and neutrinoless ($0\nu\beta\beta$) decay modes, and for five nuclei of most experimental interest. Further, using the most recent experimental limits for $0\nu\beta\beta$ decay half-lives, we provide new constraints on the light mass neutrino parameter. Finally, by separating in the half-lives formulas the factor representing the axial-vector constant to the forth, we advance suggestions on how to reduce the errors introduced in calculation by the uncertain value of this constant by exploiting the DBD data from different isotopes and/or decay modes.  


\end{abstract}

\maketitle


\section{Introduction}

The double-beta decay is a rare nuclear process intensively studied due to its potential to test nuclear structure methods and investigate beyond Standard Model (SM) physics \cite{VES12}-\cite{SC98}. According to the number and type of the released leptons there are several possible DBD modes, that can be classified in two categories. One category is that where two anti-neutrinos or two neutrinos are emitted in the final states besides the two electrons ($2\nu\beta^-\beta^- $) or two positrons ($2\nu\beta^+\beta^+ $). The double-positron decays can also be accompanied by one or two electron capture processes ($2\nu\beta^+ EC $, $2\nu ECEC $).
 These decay modes occur with lepton number conservation (LNC) and are allowed within the SM. In the other category enter decay processes similar with the above ones, but where no anti-neutrinos or neutrinos are emitted in the final states. They are generically called neutrinoless DBD processes ($0\nu\beta\beta$), so we may have $0\nu\beta^-\beta^-$, $0\nu\beta^+\beta^+$, $0\nu\beta^+ EC$ and $0\nu ECEC$ decays in this category. All these processes violate LNC, hence they are not allowed within the original framework of the SM but can appear in theories more general than the SM. The discovery of any $0\nu\beta\beta$ decay mode would firstly demonstrate the lepton number violation by two units, but would also provide us with valuable information on other beyond SM processes. From the $2\nu\beta\beta$ decay study one can get information about nuclear structure, test different nuclear methods and investigate the violation of Lorentz symmetry in the neutrino sector, while from the $0\nu\beta\beta$ decay study one can decide about the neutrino character (is it a Dirac or a Majorana particle?), one can constrain beyond SM parameters associated with different mechanisms that may contribute to this decay mode and one can get information about neutrino mass hierarchy, existence of heavy neutrinos, of right-handed components in the weak interaction currents, etc. That is why, the DBD study is a very important and timely topic.

The first step in theoretical study of the DBD process is to derive half-lives expressions and calculate the quantities therein, for each possible decay mode and for different transitions and mechanisms that may contribute to the $0\nu\beta\beta$ decay mode.  With good approximation, the DBD half-lives formulas can be written in factorized forms, as follows \cite{KI12}, \cite{SM13}: 

\begin{equation}
\label{1}
\left( T^{2\nu}_{1/2} \right)^{-1} = G^{2\nu}(E_0, Z) \times g_A^4 \times \mid m_e c^2  M^{2\nu} \mid^2  
\end{equation}

\begin{equation}
\label{2}
\left( T^{0\nu}_{1/2} \right)^{-1} = G^{0\nu}(E_0, Z) \times g_A^4 \times \mid M_l^{0\nu}\mid^2 \left( \langle\eta_{l}\rangle \right)^2 
\end{equation}
\noindent
where $G^{(2,0)\nu}$ are the PSF, $M^{(2,0)\nu}$ are the NME, for the $(2,0)\nu$ decay modes and $\langle\eta_{l}\rangle$ is a parameter related to the specific mechanism l that can contribute to the $0\nu\beta\beta$ decay. We note that the half-lives expressions from above are written such that the product of the nuclear (NME) and atomic part (PSF) is expressed in $[yr^{-1}]$. Also, we note that the axial-vector constant to the forth power is separated from the other components. Such a form of the half-lives expressions allows an easy using of the theoretical results for interpreting the DBD data and possibility to make connections between data from different decay modes and experiments in an attempt to find solutions to reduce the errors in computation related to the value of axial-vector constant which is not precisely known.
As seen, for estimating/predicting DBD lifetimes and deriving beyond SM parameters, a precise, reliable computation of both the PSF and NME is mandatory. The largest uncertainties in the DBD calculations come from the NME. They are calculated with different nuclear methods, the most currently employed being pnQRPA \cite{SC98}, \cite{ROD07}-\cite{SK01}, Shell Model \cite{CAU08}-\cite{HS10}, IBA2 \cite{BI09}-\cite{BKI13}, PHFB \cite{RAH10}, GCM with EDF\cite{RMP10}. They differ each other mainly by the choice of the model spaces and type of correlations taken into account in calculation. Each of these methods has its own advantages and drawbacks, and errors in the NME computation associated with each of them have been extensively debated in the literature over time \cite{SC98}, \cite{ROD07}-\cite{RMP10}. 	
The differences in the NME values computed with different methods may come from different sources such us i) the choice of the model space of single-particle orbitals and type of the nucleon-nucleon correlations included in calculation which are specific to different nuclear methods, ii) the nuclear structure approximations associated with the short range correlations (SRC), finite nucleon size (FNS), higher order terms in the nucleon currents (HOC), inclusion of deformation, etc., or iii) the use of input parameters whose values are not precisely known, like nuclear radius, the average energy of the virtual states in the intermediate odd-odd nuclei or the value of the axial-vector constant, $g_A$, etc. Particularly important is the value of $g_{A}$ (which can be 1.0 = quark value; 1.273 = free nucleon value; or other quenched value (0.4-0.9) because the dependence of the half-lives on this constant is strong. We note that errors coming from the different choice of values of these parameters can increase significantly the uncertainty in the half-lives computation, hence appropriate attention should also be paid to this source or errors.  
		
 On the other hand, the PSF have been considered during long time to be computed with enough accuracy \cite{SC98}, \cite{PR59}-\cite{TOM91}. However, newer calculations \cite{KI12}-\cite{SM13}, \cite{MPS15} performed with more rigorous methods, i.e. by using exact electron Dirac wave functions (w.f.) and improving the way of taking into account the finite nuclear size (FSN), electron screening effects  and more realistic form of the Coulomb potential, revealed notable differences of the PSF values as compared with older results, especially for heavier nuclei, for positron emitting and $EC$ decay modes and for transitions to excited states.  

 The errors in the PSF computation can come from i) the method of calculation of the electron w.f., namely - non-relativistic approach \cite{PR59}; -relativistic approach with approximate electron w.f. \cite{SC98}; - relativistic approach with exact electron w.f. \cite{KI12}-\cite{SM13}, \cite{MPS15}; ii) numerical accuracy both in the resolution of the Dirac equations for getting the electron radial functions and in the integration of the PSF expressions, for different decay modes.  
  
We also note that some input parameters appear both in the NME and PSF expressions, such as the axial-vector constant $g_{A}$, the nuclear radius $R_A$ $(R_A =r_0 A^{1/3})$, the value of the average energy of the virtual states in the intermediate odd-odd nucleus, used in the closure approximation, $\langle E_N \rangle$, etc. Also, when these quantities are calculated separately, different groups have used sometimes different values for these parameters. Moreover the NME and PSF have been reported in different units depending on which factors were included in their expressions, and this led sometimes to some confusion/difficulty in the theoretical predictions and interpretation of the experimental data. 

In this paper we propose a new approach of calculating the NME and PSF entering the DBD half-lives, namely to calculate directly their product, in an unique formula, instead of calculating them separately. This is actually natural, since for predicting half-lives and getting information about beyond SM physics from the DBD study, we need to know precisely the product $NME \times PSF$. The computation of the product as a whole has some advantages. Calculating its values in units of $[yr^{-1}]$ one facilitates its using in predicting and interpreting the DBD experimental data, by removing any confusion related to the units in which its components are reported when they are calculated separately. Also, the formula of the product has an unique dependence of a certain parameter, for which one takes a single value. Thus, the computation of the atomic and nuclear part of the DBD half-lives gets coherence, of which has not been paid attention to so far. 
Finally, we note that the separation of the $g_{A}^4$ factor in the half-life expressions  can also have advantages. For example, by combining experimental data and information from different DBD isotopes and/or decay modes and transitions, one can reduce the uncertainty of the calculation related to this parameter.

\section{Products of phase space factors and nuclear matrix elements}
		
	We define the products as follows: 
	\begin{equation}
	\label{3}
	 P^{2\nu} = G^{2\nu} ~ \times ~ |m_e c^2 M^{2\nu}|^2
	\end{equation}
	\begin{equation}
	\label{4}
	 P^{0\nu} = G^{0\nu}~ \times ~|M^{0\nu}_l|^2
	 \end{equation} 
	 \noindent
	 so, the half-lives expressions become:   
	
	\begin{equation}
	\label{5}
	\left( T^{2\nu} \right)^{-1} = \left( g^{2\nu}_{A, eff} \right)^4 ~ \times ~ P^{2\nu} 
	\end{equation}
	
	\begin{equation}
	\label{6}
	\left( T^{0\nu} \right)^{-1} = \left( g^{0\nu}_{A,eff} \right)^4 ~ \times ~ P^{0\nu} ~ \times ~ \langle \eta_l \rangle^2
	\end{equation}
	\noindent
	where $g_{A,eff}$ is the effective value of the $g_{A}$ constant that can be different for different nuclei and decay modes, because it can depend on nuclear medium and many-body effects. Hence, providing the products $P^{(2,0)\nu}$ in $[yr^{-1}]$ one can use them easily for predicting half-lives and/or constraining beyond SM parameters. The detailed expressions of these products read:	

i\begin{widetext}
\begin{eqnarray}
\label{7}
\it{P^{2\nu}} &=& \frac{\tilde{A}^2 \left( G \cos\theta_C \right)^4}{96\pi^7 \hbar ln 2}|M^{2\nu}|^2 \times
\int_{m_ec^2}^{Q_{\beta\beta}+m_e c^2}d\epsilon_1
\int_{m_ec^2}^{Q_{\beta\beta}+2m_ec^2-\epsilon_1}d\epsilon_2
\int_{0}^{Q_{\beta\beta}+2m_ec^2-\epsilon_1-\epsilon_2}d\omega_1 f_{11}^{(0)} \epsilon_1 \epsilon_2 \omega_1^2 \omega_2^2 (p_1c)(p_2c)  \nonumber\\[2mm]
&&
\hskip2cm \times \left[\langle K_N\rangle^2+\langle L_N\rangle^2 + \langle K_N\rangle \langle L_N\rangle\right]   \hskip7.4cm
\end{eqnarray}
\vspace{0.5cm}
\begin{equation}
\label{8}
\it{P^{0\nu}}= \frac{\left( G \cos\theta_C \right)^4 (m_e c^2)^2(\hbar) c^2} {32 \pi^5 R^2\ln 2} |M^{0\nu}|^2 \times
\int_{m_e c^2}^{Q_{\beta\beta}+m_e c^2} \epsilon_1\epsilon_2(p_1c)(p_2c)d\epsilon_1 f_{11}^{(0)}\left[\langle K_N \rangle-\langle L_N \rangle\right]^2  
\end{equation}
\end{widetext}

\noindent
where $G$ is the Fermi constant, $\theta_C$ the Cabbibo angle, $Q_{\beta\beta}$ the Q-value for the DBD, $m_e$ the electron mass,
and $\epsilon_{1,2}$ and $\omega_{1,2}$ are the electron and neutrino energies, respectively. Also: 

\begin{equation}
\label{9}
\tilde{A} = \left[\frac{1}{2}Q_{\beta\beta} +  2m_ec^2+ \langle E_N\rangle -E_I \right],
\end{equation}

\noindent
where $\langle E_N \rangle $ is an average energy of the states $E_I$ in the odd-odd intermediate nucleus that contribute to the decay. $\langle K_N \rangle$ and $\langle L_N \rangle $ are quantities that depend on the electron and neutrino energies, as well as on the energies $\langle E_N \rangle $ and $E_I$ \cite{HS84}. $f_{11}^{(0)}$ are combinations of the radial electron functions $g_k$ and $f_k$, solutions of the Dirac equations \cite{MPS15}. Finally, $M^{(2,0)\nu}$ are the NME for $2\nu$ and $0\nu$ decay modes. 
  	
For computing the products $P^{2\nu}$ and $P^{0\nu}$ we build up numerical codes taking advantage of our previous codes for computing separately the NME and PSF quantities \cite{HSB07}, \cite{MPS15}, \cite{NS14}.
The expressions of the products $P^{(2,0)\nu}$ contain factors outside the integrals stemming from the multiplication and simplification of factors that multiply separately the nuclear and kinetic parts. Also, their kinetic part (phase space factors) and the nuclear part (NME) have common input parameters as $R_A$, $\langle E_N \rangle$ and $g_A$. 

 Firstly, we refer to the $P^{2\nu}$ computation. The kinetic part is computed following the main lines of the approach developed in our previous works from refs. \cite{SM13}-\cite{MPS15} and here we shortly review the main ingredients of the code and computation. We first use a subroutine where the electron wave functions are got as radial solutions ($g_k$ and $f_k$) with appropriate asymptotic behavior of the Dirac equations with a Coulomb-type potential, and including the finite nuclear size and electron screening effects. The Coulomb-type potential is obtained from a realistic proton density in the daughter nucleus. For getting the single particle densities inside the daughter nucleus, we solve the Schrodinger equation for a spherical Woods-Saxon potential with spin orbit and Coulomb terms \cite{SM13}-\cite{MPS15}. Then, the PSF part of the code is completed by performing the integrals over the electron phase factors build up with the Dirac radial functions. The code has an improved numerical accuracy for finding the electron w. f. and a better interpolation procedure for integrating the PSF final expressions, especially at low electron energies. For the NME part we use a code similar to that from ref. \cite{HSB07} for computing the double Gamow-Teller transitions, using the following effective nucleon-nucleon interactions: GXPF1A \cite{HON04} for $^{48}Ca$, JUN45 \cite{HON09} for $^{76}Ge$ and $^{82}Se$ and gcn50:82 \cite{CAU09} for $^{130}Te$ and $^{136}Xe$.\\
The  values for the products $P^{2\nu}$ are presented in the third column of the Table 1 for five nuclei of experimental interest. With the values of $g^{2\nu}_{A, eff}$ written in the forth column of the table as first entries, we reproduced the most recent measured half-lives found in literature, which are displayed in the second column. In the forth column we also show the $g^{2\nu}_{A,exp}$ values taken from Refs. \cite{IWA16}-\cite{CNP12}, which were obtained by comparing the theoretical B(GT) strengths with the  experimental ones extracted from charge-exchange reactions. In the last column we present the difference in percentage between the $g_{A,eff}$ values obtained within our calculations to reproduce the experimental DBD half-lives and those obtained by fitting the B(GT) experimental data, estimated in percentage, $\epsilon =\left(g^{2\nu}-g^{2\nu}_{A,eff}\right)/g^{2\nu}$. As seen, the two sets of values are close to each other, the smallest differences being in the case of $^{76}Ge$ and $^{48}Ca$ nuclei.

\begin{table}

\begin{center}
	\caption{ \label{tab1} Results for $2\nu\beta\beta$ decay mode}
	\footnotesize
	\begin{tabular*}{80mm}{c@{\extracolsep{\fill}}ccccc}
		\toprule Nucleus & $T_{1/2}^{2\nu}[yr]$ & $P^{2\nu}[yr^{-1}]$ & $g^{2\nu}_{A,eff}/g^{2\nu}_{exp}$& $\epsilon [\%]$ \\
		\hline
		$^{48}$Ca & $6.40\times 10^{19}$ \cite{NEMO16} & $123.81\times10^{-21}$ & 0.65/0.71\cite{IWA16}& 8.45 \\
		$^{76}$Ge & $1.92 \times 10^{21}$ \cite{PAT16} & $5.16 \times 10^{-21}$ & 0.56/0.60\cite{CNP12} & 6.60 \\
		$^{82}$Se & $0.92 \times 10^{20}$ \cite{PAT16} & $186.62 \times 10^{-21}$  & 0.49/0.60\cite{CNP12} & 18.33\\
		$^{130}$Te & $8.20 \times 10^{20}$ \cite{CUORE17} & $25.26 \times 10^{-21}$ & 0.47/0.57\cite{CNP12} & 17.33\\
		$^{136}$Xe & $2.16 \times 10^{21}$ \cite{NEMO16} & $20.30 \times 10^{-21}$ &0.39/0.45\cite{CNP12} & 13.33\\
	\end{tabular*}
\end{center}
\end{table}

Then, we calculated the $P^{0\nu}$ products in the case of the light Majorana neutrino exchange mechanism, with $\langle\eta_l\rangle = \langle m_{\nu}\rangle/{m_e}$ and the light neutrino parameter defined as:

\begin{equation}
\label{10}
\langle m_{\nu} \rangle = \mid \Sigma_{k=1}^3 U_{ek}^2 m_k\mid 
\end{equation} 
\noindent
where $U_{ek}$ are the first row elements of the Pontecorvo-Maki-Nakagawa-Sakata neutrino mixing matrix and $m_k$ are the light neutrino masses \cite{KL13}. 
The expression of the nuclear matrix elements can be written as a sum of Gamow-Teller ($GT$), Fermi ($F$) and tensor ($T$) components \cite{SIM09}, \cite{NS14}:

\begin{equation}
\label{11}
M^{0 \nu}=M^{0 \nu}_{GT}-\left( \frac{g_V}{g_A} \right)^2  M^{0 \nu}_F + M^{0 \nu}_T \ ,
\end{equation}
\noindent
where $M^{0 \nu}_{GT}$, $M^{0 \nu}_F$ and $M^{0 \nu}_T$ are these components. These are defined as follows:
\begin{equation}
\label{12}
M_\alpha^{0\nu} = \sum_{m,n} \left< 0^+_f\| \tau_{-m} \tau_{-n}O^\alpha_{mn}\|0^+_i \right> \ ,
\end{equation}
\noindent
$O^\alpha_{mn}$ are transition operators ($\alpha=GT,F,T$) and the summation is over all the nucleon states.
Correspondingly, the two-body transition operators $O^{\alpha}_{12}$ can be expressed in a factorized form as \cite{NS14}:
\begin{equation}
\label{13}
O^{\alpha}_{12} = N_{\alpha}  S_{\alpha}^{(k)} \cdot  \left[R_{\alpha}^{(k_r)}\times C_{\alpha}^{(k_c)}\right]^{(k)}
\end{equation}
\noindent
where $N_\alpha$ is a numerical factor including the coupling constants, and  $S_\alpha$, $R_\alpha$ and $C_\alpha$ are operators acting on the spin, relative and center-of-mass wave functions of the two-particle states, respectively.
Thus, the calculation of the matrix elements of these operators can be decomposed into products of reduced matrix elements within the two subspaces \cite{HS10}. The expressions of the two-body transition operators are:
\begin{eqnarray}
\label{14}
O_{12}^{GT} = \sigma _1 \cdot \sigma _2 H(r), ~~ O_{12}^{F} = H(r) \hskip3cm &&	\nonumber\\[2mm]  O_{12}^{T} = \sqrt{\frac{2}{3}} \left [ \sigma _1 \times \sigma _2 \right ]^2 \cdot \frac{r}{R} H(r) C^{(2)}(\hat r) \hskip2.5cm
\end{eqnarray}

The $O^{\alpha}_{12}$ operators contain three components, namely, the spin, center-of-mass and radial ones, and the expectation values of the first two components can be easily managed.  The radial part is the most difficult to be calculated because it contains neutrino potentials written in different approximations, and the expectation values are double integrals over them. Also, short-range correlations and finite nucleon size corrections are introduced in this part of computation. 
The neutrino potentials depend weakly on the intermediate states, and are defined by integrals over momentum carried by the virtual neutrino exchanged between the two nucleons \cite{SIM09}. They include Fermi (F), Gamow-Teller (GT) and tensor (T) components: 
\begin{eqnarray}
\label{15}
H(r) &=& \frac{2R}{\pi} \int^\infty_0 \frac{q dq}{q + \langle E_N \rangle} \nonumber \\ [2mm]
&&
\times  \left[ j_0(qr) \left(h_F(q) + h_{GT}\right) + j_2(qr) h_T\right] \hskip1cm
\end{eqnarray}\\

\noindent 
where $R=r_0 A^{1/3}$ fm, with $r_0=1.2fm$, $j_{0,2}(qr)$ are the spherical Bessel functions and the integrals are over the neutrino exchange momentum ${\it q}$.
In our calculations we use the closure approximation and $\langle E_N\rangle$, as mentioned above, represents the average energy of the virtual states in the intermediate odd-odd nucleus included in the description of the decay. Also, we note that the factor $2R$ is canceled by the similar one from the denominator of the PSF expression, so the $P^{0\nu}$ does not depend on the nuclear radius.   The expressions of neutrino potentials $h_{F,GT,T}$ can be found in many references (see for example \cite{SIM09}). 
These expressions include FNS effects taken into account through vector and axial-vector form factors, $G_V$ and $G_A$ \cite{SIM09}.

\begin{equation}
\label{16}
G_A \left(q^2 \right) = g_A \left( \frac{\Lambda^2_A}{\Lambda^2_A+q^2} \right)^2, \ G_V \left( q^2 \right) = g_V \left( \frac{\Lambda^2_V}{\Lambda^2_V+q^2} \right)^2
\end{equation} 
We take the following values for the vector and axial vectors form factors: $\Lambda_V=850 MeV$ and $\Lambda_A=1086 MeV$ \cite{AEE08}.

For computing the radial matrix elements \\ $\langle nl|H_{\alpha}|n'l'\rangle$ we use
the harmonic oscillator $HO$ wave functions  $\psi_{nl}(lr)$ and $\psi_{n^\prime l^\prime}(r)$ corrected by a factor \\ $[1 + f(r)]$, which takes into account the SRC effects induced by the nuclear interaction \cite{NS14}:
\begin{equation}
\label{17}
\psi_{nl}(r) \rightarrow \left[ 1+f(r) \right] \psi_{nl}(r) \ 
\end{equation}
For the correlation function we take the functional form
\begin{equation}
\label{18}
f(r) = - c \cdot e^{-ar^2} \left( 1-br^2 \right) \ 
\end{equation}
For the $a$, $b$ and $c$ constants we use the parametrization used in ref. \cite{ccm}. \\

Including HOC and FNS effects the radial matrix elements of the neutrino potentials become:
\begin{eqnarray}
\label{19}
\langle nl || H_\alpha(r) || n^{\prime} l^{\prime} \rangle &=& \int^\infty_0 r^2 dr \psi_{nl}(r) \psi_{n^\prime l^\prime}(r)\left[ 1+f(r) \right]^2 \nonumber \\[2mm]
&&
\times  \int^\infty_0 q^2 dq V_{\alpha} (q) j_0 (qr) 
\end{eqnarray}

\noindent
We note that in the case of $P^{0\nu}$ products, the axial-vector constant enters also the expressions of the neutrino potentials, in addition to the factor $g_{A,eff}^4$, so the half-life expression for the $0\nu\beta\beta$ decay, Eq. (5), contains a "double" dependence on this constant. Of course, for coherence, the same values of  the $g_{A,eff}$ constant should be taken in both places, i.e. both in the ${\it P^{0\nu}}$ and in the half-life computation. We note that these values may differ from the values of this constant used in the $2\nu\beta\beta$ decay mode. Because we do not know until now what is the correct value of $g_{A.eff}$ for $0\nu\beta\beta$ decay mode, we calculated the $P^{0\nu}$ products for the free nucleon value (1.273). Being an input parameter, the $P^{0\nu}$ values can be easily computed for other effective values of this constant. The obtained values of the products $P^{0\nu}$, in $[yr^{-1}]$ units, are presented in the third column of the Table 2. 
\begin{table}
\begin{center}
	\caption{ \label{tab2} Results for the  $0\nu\beta\beta$ decay mode}
	\footnotesize
	\begin{tabular*}{80mm}{c@{\extracolsep{\fill}}ccc}
		\toprule Nucleus &  $T_{1/2}^{0\nu} [yr]$ & $P_\nu^{0\nu} [yr^{-1}]$ & $\langle m_\nu\rangle$[eV] \\
		\hline
		$^{48}$Ca & $> 2.0 \times 10^{22}$ \cite{NEMO16} & $7.30 \times 10^{-15}$ & < 26.49 \\
		$^{76}$Ge &$> 8.0 \times 10^{25}$ \cite{GERDA18} & $9.95 \times 10^{-15}$ & < 0.29 \\
		$^{82}$Se& $> 3.6\times10^{23}$ \cite{TRE11} & $34.45 \times 10^{-15}$ & < 2.87 \\
		$^{130}$Te &$> 4.0\times10^{24}$ \cite{CUORE15} & $71.45 \times 10^{-15}$ & < 0.59 \\
		$^{136}$Xe & $> 1.8\times10^{25}$ \cite{EXO18} & $71.01 \times 10^{-15}$ & < 0.28\\
	\end{tabular*}
\end{center}
\end{table}

At this point it is worth to mention that the values of the products $P^{(2,0)nu}$ from Tables 1 and 2, obtained with the approach described here, are very close to the values that we obtained when we computed these products by multiplying the values of NME and PSF calculated separately, as one should be. This is understandable because in their calculation by the two methods we used the same values of the input parameters and the same nuclear approximations and parametrizations. The small differences come from the numerical precision of the numerical codes, we used. We emphasize, however, that the importance of our current approach is that it can eliminate the incoherence of using NME and PSF values calculated separately with different values for common nuclear parameters, which can introduce significant errors in the evaluation of $NME \times PSF$ product as a whole. The errors in the evaluation of these products can indeed be significant if one takes different values of $g_A$ in the computation of $NME$ and $PSF$ and if these values are not the same with the value used in the $g_A^4$ factor.  For example, the errors introduced in the NME computation by the use of a quenched (1.0) or an unquenched (1.27) value of  $g_A$ were analyzed in Ref. \cite{NS14} for $^{48}Ca$,  $^{76}Ge$ and  $^{82}Se$ nuclei, and found to be  within 10-14\% (without the factor $g_A^4$).  The use of different values for the other (common) parameters involved in calculations as the nuclear radius, $<E_N>$, etc. can bring additional uncertainties of the same order. The errors can be amplified by the use of different values of these parameters in the PSF computation. So, there may be relevant errors in calculating the products $NME \times PSF$ when the $NME$ and $PSF$ values are taken from separate calculations reported in literature.

Then, we revise the limits of the light neutrino mass parameter $\langle m_{\nu}\rangle$ using our calculations and the most recent experimental limits reported for the $0\nu\beta\beta$ decay half-lives. These results are presented in the last column of the Table 2.  One observes that presently the most stringent constraints on this parameter come from the nuclei $^{76}Ge$ and $^{136}Xe$, due to both the experimental results (the lowest limits measured at present for the $0\nu\beta\beta$ decay half-lives \cite{GERDA18}, \cite{EXO18}) and to accurate theoretical calculations. An important issue in this case remains the use of a correct value for the $g_A$ constant. As far as this value is still unknown, for accurate half-lives predictions and constrains of beyond SM parameters, the goal is to reduce the errors associated with this constant. One suggestion is to use information from different decay modes and/or from DBD experiments on different nuclei. For example, for a particular nucleus the ratio of the $2\nu$ and $0\nu$ half-lives expressions reads:

\begin{equation}	
\label{20}
\left(\frac{T^{2\nu}}{T^{0\nu}}\right) = \left( \frac{g^{0\nu}_{A,eff}}{g^{2\nu}_{A,eff}} \right)^4 \times \frac{P^{0\nu}}{P^{2\nu}} \times \langle \eta_l \rangle^2 	
\end{equation}
\noindent
As seen from the above formula, any information that we can get about the relative magnitude of the $g_{A, eff}$ values for the $2\nu$ and $0\nu$ decays in the same nucleus, can be exploited to improve the constraints on the neutrino mass parameter, when improved calculations of $P^{(2,0)}$ are available.  
Also, referring to two different nuclei, denoted with $m$ and $n$, the ratio of their half-lives reads: 

For    $2\nu\beta\beta$ decay mode:	
\begin{equation}	
\label{21}
\left( T^{2\nu} \right)_n = \left( \frac{g^{2\nu}_{A,eff}(m)}{g^{2\nu}_{A,eff}(n)} \right)^4 \times \left( \frac{P^{2\nu}_m}{P^{2\nu}_n} \right) \times \left( T^{2\nu} \right)_m	
\end{equation}
\noindent

For $0\nu\beta\beta$ decay mode we have:

\begin{equation}
\label{22}
\left( T^{0\nu} \right)_n = \left( \frac{g^{0\nu}_{A, eff}(m)}{g^{0\nu}_{A, eff}(n)} \right)^4 \times \left( \frac{P^{0\nu}_m}{P^{0\nu}_n} \right) \times \left( T^{0\nu} \right)_m  
\end{equation}
\noindent
As seen, one can deduce $g^{2\nu}_{A,eff}$ for one particular nucleus if one knows with (more) precision the value of this constant for another nucleus, using the experimental half-lives and the calculated $P^{2\nu}$ for both nuclei. For example, one can take advantage of the possible experimental determination of this parameter for some particular isotopes, as it was recently proposed in Ref. \cite{SIM18}. Similar considerations, i.e. the exploitation of data from several experiments are valid for predicting $2\nu\beta\beta$ decay half-lives for a nucleus that was not yet measured, if we have accurate data for another nucleus and good estimations of the $g^{2\nu}_{A,eff}$ value from other (non-DBD) experimental data. For such predictions information on DBD half-lives not yet measured obtained from empirical formulas, as is proposed in Ref. \cite{YZR2014}, are valuable.   

Similarly, for the $0\nu\beta\beta$ decay one can deduce more information about the effective value of   $g^{0\nu}_A$ for a particular nucleus if we know this value for another. For example, we might know 
$g^{0\nu}_{A,eff}$ with more precision in the case of nuclei where single state dominance (SSD) approximation is valid, where the half-life can be computed with reasonable precision by taking into account only one state in the intermediate odd-odd nucleus (for example $^{100}$Mo case), and where $g^{2\nu}_{A,eff}$ and $g^{0\nu}_{A,eff}$ might have close values.

\section{Conclusions}

We proposed a new approach of calculating the NME and PSF for DBD, by computing directly their product. The product as a whole can be computed more consistently, having an unique dependence of some parameters entering previously separately the NME and PSF expressions and taking thus single values for them. The values of the product are given in the same units as $T_{1/2}^{-1}$ (i.e. [$yr^{-1}]$) removing any possible confusion in using the theoretical calculations for interpreting the DBD data. The new codes of calculating the $NME \times PSF$ products include improved routines used in our previous papers for the separately computation of these two quantities. We provide values of these products for $2\nu$ and $0\nu$ DBD modes for five nuclei of most experimental interest.
Then, using our calculations and the newest half-life values for the $0\nu\beta\beta$ decays reported in literature, we revise the upper limits for the light mass neutrino parameter. In the half-lives formulas we separate the strong dependence on the axial-vector constant, i.e. the factor $g_A^4$, that bring a large uncertainty in calculation and suggest some ways to reduce/avoid the errors related to the uncertain value of this constant. This could be done by using ratios of $g^{(2,0)\nu}_{A, eff}$ and $P^{(2,0)\nu}$ (instead of their individual values) and exploiting data on the same nucleus but for different decay modes and/or DBD data from experiments on different nuclei, including the possibility that this constant to be determined experimentally for some particular isotopes. We hope our work will be a forward step for more consistent DBD calculations and which will be easier used in predicting and interpreting the experimental data.


\end{document}